\pdfoutput=1
\documentclass[fleqn,10pt]{wlscirep}
\title{Longitudinal spin Seebeck coefficient: heat flux vs. temperature difference method}

\author[1,*]{A. Sola}
\author[2]{P. Bougiatioti}
\author[1]{M. Kuepferling}
\author[2]{D. Meier}
\author[2]{G. Reiss}
\author[1]{M. Pasquale}
\author[2,3]{T. Kuschel}
\author[1]{V. Basso}
\affil[1]{Istituto Nazionale di Ricerca Metrologica, Strada delle Cacce 91, 10135, Turin, Italy}
\affil[2]{Center for Spinelectronic Materials and Devices, Department of Physics, Bielefeld University, Universit{\"a}tsstrasse 25, 33615 Bielefeld, Germany}
\affil[3]{Physics of Nanodevices, Zernike Institute for Advanced Materials, University of Groningen, Nijenborgh 4, 9747 AG Groningen, The Netherlands}

\affil[*]{a.sola@inrim.it}

\begin{abstract}
The determination of the longitudinal spin Seebeck effect (LSSE) coefficient is currently plagued by a large uncertainty due to the poor reproducibility of the experimental conditions used in its measurement. In this work we present a detailed analysis of two different methods used for the determination of the LSSE coefficient. We have performed LSSE experiments in different laboratories, by using different setups and employing both the temperature difference method and the heat flux method. We found that the lack of reproducibility can be mainly attributed to the thermal contact resistance between the sample and the thermal baths which generate the temperature gradient. Due to the variation of the thermal resistance, we found that the scaling of the LSSE voltage to the heat flux through the sample rather than to the temperature difference across the sample greatly reduces the uncertainty. The characteristics of a single YIG/Pt LSSE device obtained with two different setups was $(1.143\pm0.007)\cdot 10^{-7}$ Vm/W and $(1.101\pm0.015)\cdot 10^{-7}$ Vm/W with the heat flux method and $(2.313\pm0.017)\cdot 10^{-7}$ V/K and $(4.956\pm0.005)\cdot 10^{-7}$ V/K with the temperature difference method. This shows that systematic errors can be considerably reduced with the heat flux method.
\end{abstract}
\begin{document}

\flushbottom
\maketitle
%
%
\thispagestyle{empty}

\section*{Introduction}

The interactions between charge carriers and heat currents is a topic of great interest both for fundamental research and technological applications such as thermoelectric power generators\cite{heremans2014thermoelectricity}. 
This class of devices exploits the Seebeck effect which allows the conversion of heat into electricity through the electric field which occurs in a junction of two different materials under a thermal gradient\cite{disalvo1999thermoelectric}. 
However, the low conversion efficiency of the Seebeck effect and the difficulty to build thin devices has limited the industrial application as heat harvesters\cite{siegel2014}.

Novel devices based on the spin Seebeck effect (SSE)\cite{uchida2008observation} may overcome these limits\cite{siegel2014,kirihara2016flexible,uchida2016thermoelectric}, thanks to a spin-current-mediated conversion from thermal flux into thermovoltage; a phenomenon which is part of the research field of spin-caloritronics\cite{bauer2012spin,boona2014spin}, which studies the interactions between spin and heat currents in magnetic materials.
The SSE refers to the rising of a spin current in a magnetic material as a consequence of a thermal gradient. The thermally generated spin current is converted into a detectable charge accumulation in a non-magnetic heavy metal material with a high spin-orbit coupling adjacent to the magnet by the inverse spin Hall effect (ISHE)\cite{Saitoh2006}.
The SSE allows a simpler device geometry because it occurs in a layered structure, while conventional thermoelectric generators are based on doped semiconductors arranged in an array of junctions.
The SSE is usually measured in the longitudinal configuration (LSSE)\cite{uchida2010observation} where the temperature gradient is applied perpendicularly to the sample plane and the applied magnetic field. In this configuration other thermoelectric contributions to the SSE result negligible\cite{kikkawa2013longitudinal,kikkawa2013separation,schmid2013transverse,meier2013influence,huang2011intrinsic,avery2012observation,meier2015longitudinal}.

A typical LSSE device is a ferrimagnetic insulating layer, e.g. Y$_3$Fe$_5$O$_{12}$ yttrium iron garnet (YIG) of a proper thickness\cite{kehlberger2015length}, covered by a thin film of a strong spin-orbit coupling material, e.g. Platinum; the thin YIG layer is grown on a much thicker non magnetic oxide substrate (i.e. YAG Yttrium Aluminum Garnet or GGG Gadolium Gallium Garnet).
Given large uncertainties of the experimental results for the YIG-Pt bilayer\cite{uchida2016thermoelectric}, especially regarding the values of the LSSE coefficients, a quantitative description of the characteristics of a LSSE device is still lacking. 
In the literature one may find different experimental setups using different ways of determining and generating $\nabla T$ such as Joule heating in an external heater\cite{uchida2008observation,meier2013thermally}, laser heating\cite{weiler2012local}, Peltier heating\cite{uchida2010observation,schmid2013transverse,shestakov2015dependence}, current-induced heating in the sample\cite{schreier2013current}, heating with electric contact needles\cite{meier2013influence,meier2015longitudinal}, on-chip heater devices\cite{wu2015spin,vlietstra2014simultaneous} and rotatable thermal gradients\cite{reimer2016quantitative}.
However, it is clear that the LSSE coefficient must not depend on the choice of set-up or measurement method but only on the material and device geometry. In this way, the results can be used to improve theoretical models\cite{Xiao2010,ohnuma2013spin,adachi2010gigantic,basso2015non} and to boost technology transfer.
For example, features like the energy conversion efficiency of a LSSE device and its relations with characteristics such as chemical or physical properties (e.g. conductivity, bandgab energy, etc.) have just been started to be investigated\cite{bou2017}. The first step of the research in this strategic field and towards the quantitative description of LSSE is the establishment of a reproducible measurement procedure.

The LSSE coefficient is defined as $S_{\textrm{LSSE}}=-E_{\textrm{ISHE}}/\nabla T$\cite{rezende2014magnon}.
$E_{\textrm{ISHE}}$ is the electric field induced by the ISHE, obtained as $E_{\textrm{ISHE}}=V_{\textrm{ISHE}}/L$ where $V_{\textrm{ISHE}}$ is the measured voltage and $L$ is the distance between the electrical contacts on the ISHE film. The thermal gradient $\nabla T$ is obtained as $\nabla T= \Delta T/L_{z}$ where $\Delta T$ is the temperature difference between the two surfaces of the sample and $L_{z}$ its thickness.
The determination of a LSSE coefficient requires the simultaneous measurement of the $V_{\textrm{ISHE}}$ and the thermal gradient $\nabla T$.
While the measurement of $V_{\textrm{ISHE}}$ can be performed with high accuracy, $\nabla T$ is not easily accessible by direct measurements. Therefore, we will show a comparison between two different experimental procedures for the determination of $\nabla T$.

\begin{figure}[ht]
	\centering
	\includegraphics[width=16cm]{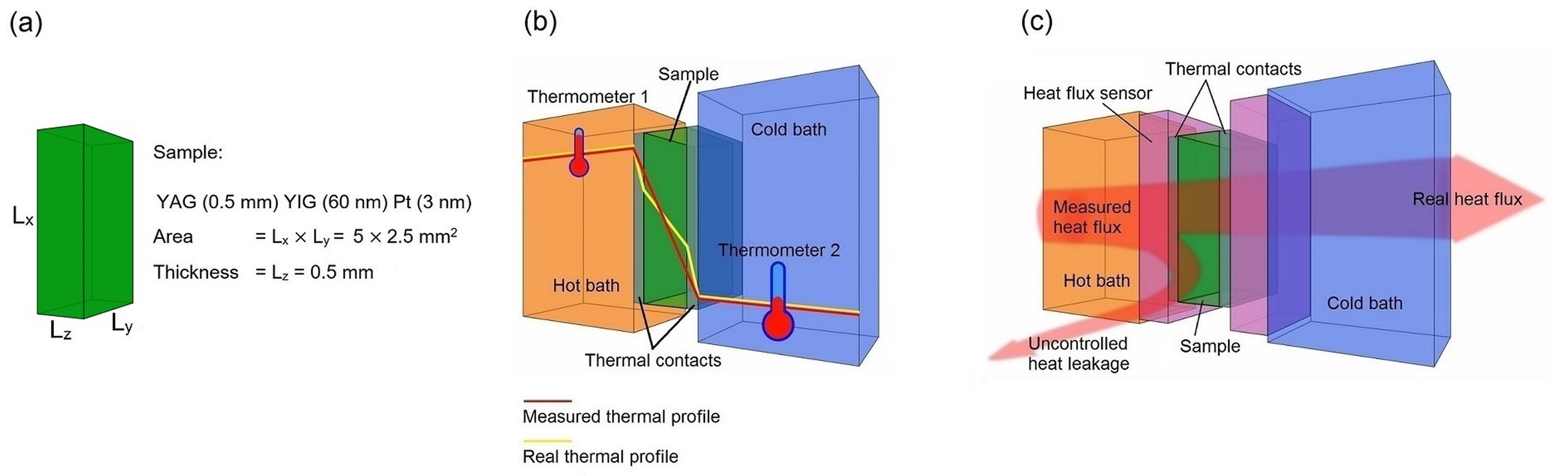}
	\caption{\label{Figure-1}
		(a) Lateral dimensions of the YAG/YIG/Pt sample adopted for the comparison.
		(b) Schematic representation of the direct temperature measurement; thermometers are placed on the hot and cold baths each one at thermal equilibrium. The red line is the assumed linear temperature profile between the two heat baths. The yellow line represents the real temperature profile regarding thermal resistances at the interfaces.
		(c) Schematic representation of the heat flux measurement; the approximation of the measured heat flux with the real quantity which flows into the sample depends on the amount of lost heat (small arrow). Dimensions of the structure, slopes of temperature profiles and proportions of heat fluxes are only qualitative and not drawn to scale.	
	}
\end{figure}

The first method that we analyze is based on the measurement of the temperature difference between the thermal baths in contact with the LSSE sample, as shown in Figure \ref{Figure-1}(b); this method was used by a majority of the research groups since the first observation of LSSE\cite{uchida2010observation}.
The method uses thermocouples for the measurement of the temperature of the thermal baths, which are large metallic heat conductors, so that it is possible to assume thermal equilibrium.
In order to estimate the temperature difference across the sample it is necessary to know the thicknesses and the thermal conductivities of all the layers found between the two thermal baths.
A common assumption is to consider a linear gradient, as represented by the red line in Figure \ref{Figure-1}(b).
This assumption is reasonable for a typical LSSE device since the thermal conductivity of the YIG layer and the substrate are similar and the heavy metal layer is thin and a good thermal conductor (Figure \ref{Figure-1}(a)).
However, this assumption neglects the temperature drop between the thermal baths and the sample due to the thermal resistance of the contacts. An example of the temperature profile in this case is represented by the yellow line in Figure \ref{Figure-1}(b) and the thermal gradients represented by the two lines may differ considerably, leading to large uncertainties in the determination of $\Delta T$\cite{sola2015evaluation}.

The second method is a possible measurement procedure developed in order to neglect the contribution of the thermal resistance of the contacts\cite{sola2015evaluation} as shown in Figure \ref{Figure-1}(c). 
This method allows the LSSE characterization as a function of the heat $Q$ flowing through the cross section of the sample $A$\cite{prakash2016spin}. This measurement is performed using calibrated heat flux sensors.
The heat flux method can be considered equivalent to the $\Delta T$ method if the thermal conductivity $k=(\frac{Q}{A})/(\frac{\Delta T}{L_{z}})$ of the magnetic material is known.
From the heat flux measurement it is possible to obtain a characterization of the LSSE sample in terms of $\frac{V_{\textrm{ISHE}}}{L}/\frac{Q}{A}$, which multiplied by the thermal conductivity $k$ gives the LSSE coefficient $S_{\textrm{LSSE}}=(\frac{V_{\textrm{ISHE}}}{L})/(\frac{\Delta T}{L_{z}})=(-E_{\textrm{ISHE}})/(\nabla T)$.
The accuracy of the heat flux method is determined by the assumption that the LSSE sample and the heat flux sensor are series elements with a negligible heat leakage from the thermal circuit.
The ohmic analogue of the heat flux method is a current measurement in a circuit with unknown resistors, which in this case are the thermal contacts at the interfaces between the sample and the heat baths.
The measured heat flux can be underestimated, as described in the sketch of Figure \ref{Figure-1}(c) if there is a heat leakage due to the electrical connections of the sample\cite{meier2013influence}, or overestimated if an uncontrolled loss takes place at the level of the heat flux sensor.

\section*{Results}

\begin{figure}[ht]
	\centering
	\includegraphics[width=11.5cm]{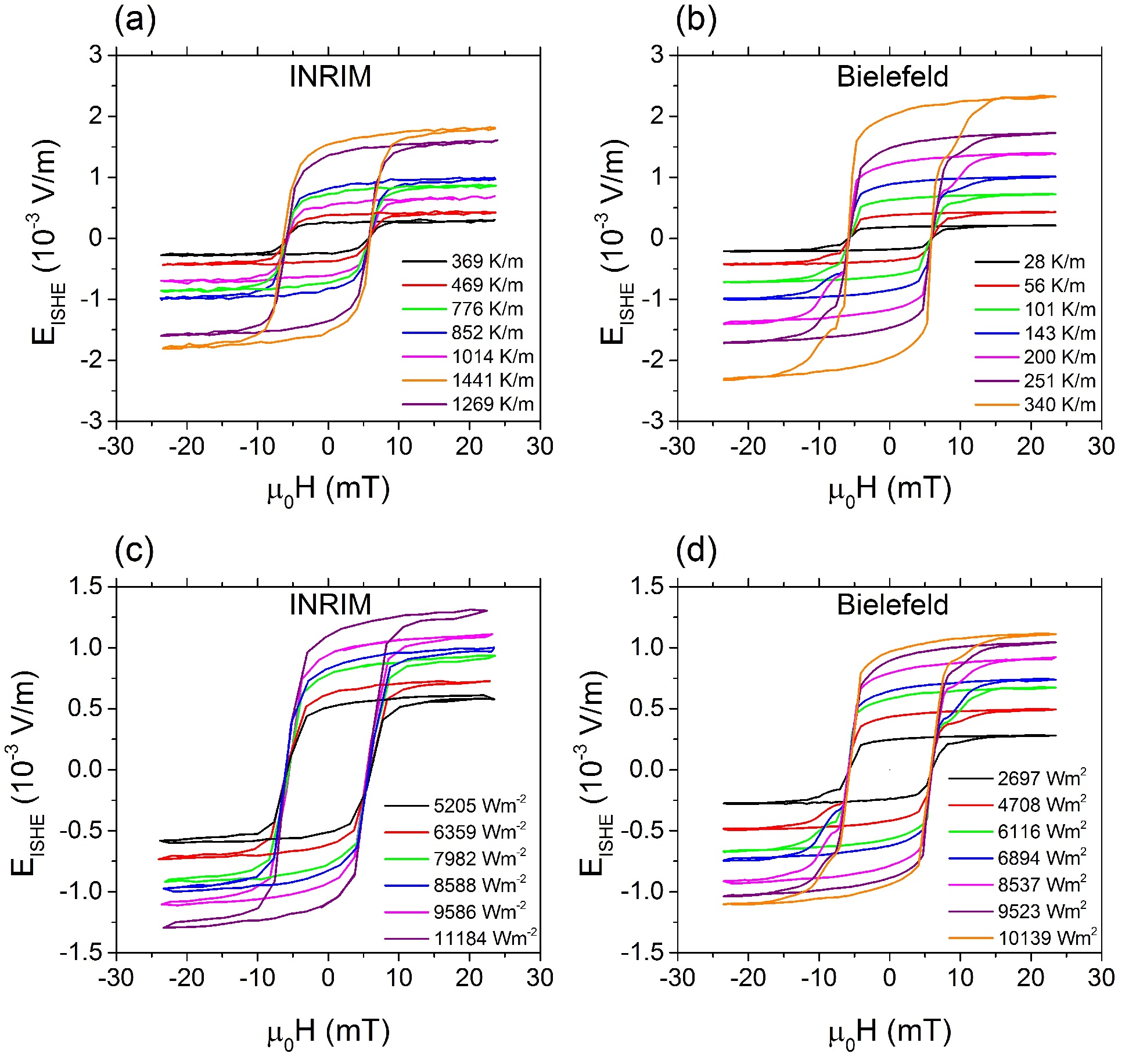}
	\caption{\label{Figure-2}
		LSSE electric field as a function of the applied magnetic field obtained (a) by INRIM using the $\Delta T$ method (b) by Bielefeld University using the $\Delta T$ method. The same measurement is repeated with the heat flux method (c) by INRIM (d) and by Bielefeld University.
	}
\end{figure}

In order to investigate the reproducibility of the measurements obtained with the two aforementioned methods, a comparison between two $\Delta T$-based measurement setups at INRIM and Bielefeld University was performed. Then both setups were modified for the measurement using the heat flux method. It was possible to estimate the systematic error using each method by measuring the characteristic of the same LSSE sample and more precisely by measuring the LSSE voltage during a saturating magnetization cycle under a given thermal gradient.
An example of measurements performed both with the $\Delta T$ and the heat flux methods is reported in Figure \ref{Figure-2}. Even though we used the same LSSE sample for all the measurements, there are some small differences between the magnetization reversal processes detected in INRIM and in Bielefeld. These are probably due to a small uncontrolled difference in alignment between the LSSE sample and the external magnetic field. 
In order to correct the effects of this misalignment on the values of the coercive fields reported in Figure \ref{Figure-2}, we performed an additional magnetization measurement by means of a vibrating sample magnetometer (VSM). This system allows a more accurate determination of the mutual positions of sample, the gaussmeter Hall probe and the magnetic field, with respect to the system that we used for the LSSE measurements. 
However, only the value of the LSSE electric field at the saturation magnetic field ($\textgreater$20 mT) is considered for the evaluation of the LSSE coefficient. 

The LSSE electric fields are obtained by the LSSE voltage drops divided by the distance between the two electrodes on the Pt-surface of the LSSE sample. There is a small difference between the noise level in INRIM and in Bielefeld that is due to the different fabrication of electrical contacts: in INRIM we used a drop of silver paste while in Bielefeld we used wire-bonding.
Using the $\Delta T$ method, each loop in magnetic field is recorded at a given value of thermal gradient that is the temperature difference recorded by the two thermocouples over the thickness of the LSSE sample.
For the heat flux method, we record the LSSE electric field at given values of heat flux obtained from the output of the calibrated Peltier sensor in Watts over the surface of the sample in square meters.

\begin{figure}[ht]
	\centering
	\includegraphics[width=17cm]{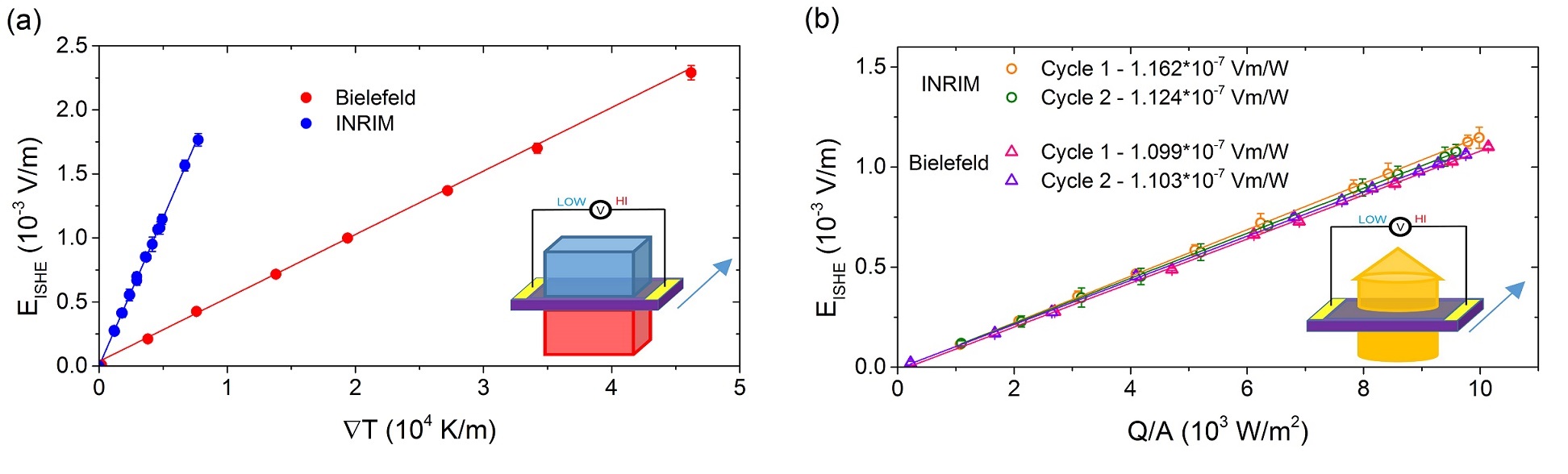}
	\caption{\label{Figure-3}
		(a) LSSE driven electric field as a function of thermal gradient obtained with $\Delta T$ method at Bielefeld University and INRIM.
		(b) LSSE electric field as a function of the heat flux measured at Bielefeld University and INRIM.
		The two drawings in the insets show the signs of the thermal gradient, the heat flux, the LSSE driven electric fields and the applied magnetic fields whose value is 20 mT for all the data.
	}
\end{figure}

The comparison between the two methods performed by the two groups are shown in Figure \ref{Figure-3}. The error bars reported in each graph represent the propagation of uncertainties: for what concerns the vertical axis of both Figure \ref{Figure-3}(a) and (b), the magnitude of the errors includes the uncertainty on the measurement of the distance between the two contacts and the uncertainty on the $V_{\textrm{ISHE}}$ voltages. We obtain this last value from the standard deviation of the set of voltages $V_{\textrm{ISHE}}$ that we record at the saturation magnetic field. The error bars on the x-axis of Figure \ref{Figure-3}(a) originate from the standard deviation of set of voltage measurements between the pair of thermocouples that we record in a steady state regime, i.e. the equilibrium condition at the two thermal baths. We apply the same procedure on the output voltages of the Peltier sensors for the evaluation of the error bars on the x-axis of Figure \ref{Figure-3}(b), which include also the uncertainties on the area measurement and the sensitivity of the Peltier sensor. The amplitudes of the error bars on the x-axis are too small to be resolved in this Figure.
The characteristics of the LSSE sample obtained from the heat flux method (Figure \ref{Figure-3}(b)) are exhibiting an average value equal to $(1.143\pm0.007)\cdot10^{-7}$ Vm/W and $(1.101\pm0.015)\cdot10^{-7}$ Vm/W from INRIM and Bielefeld University, respectively.

\begin{figure}[ht]
	\centering
	\includegraphics[width=14cm]{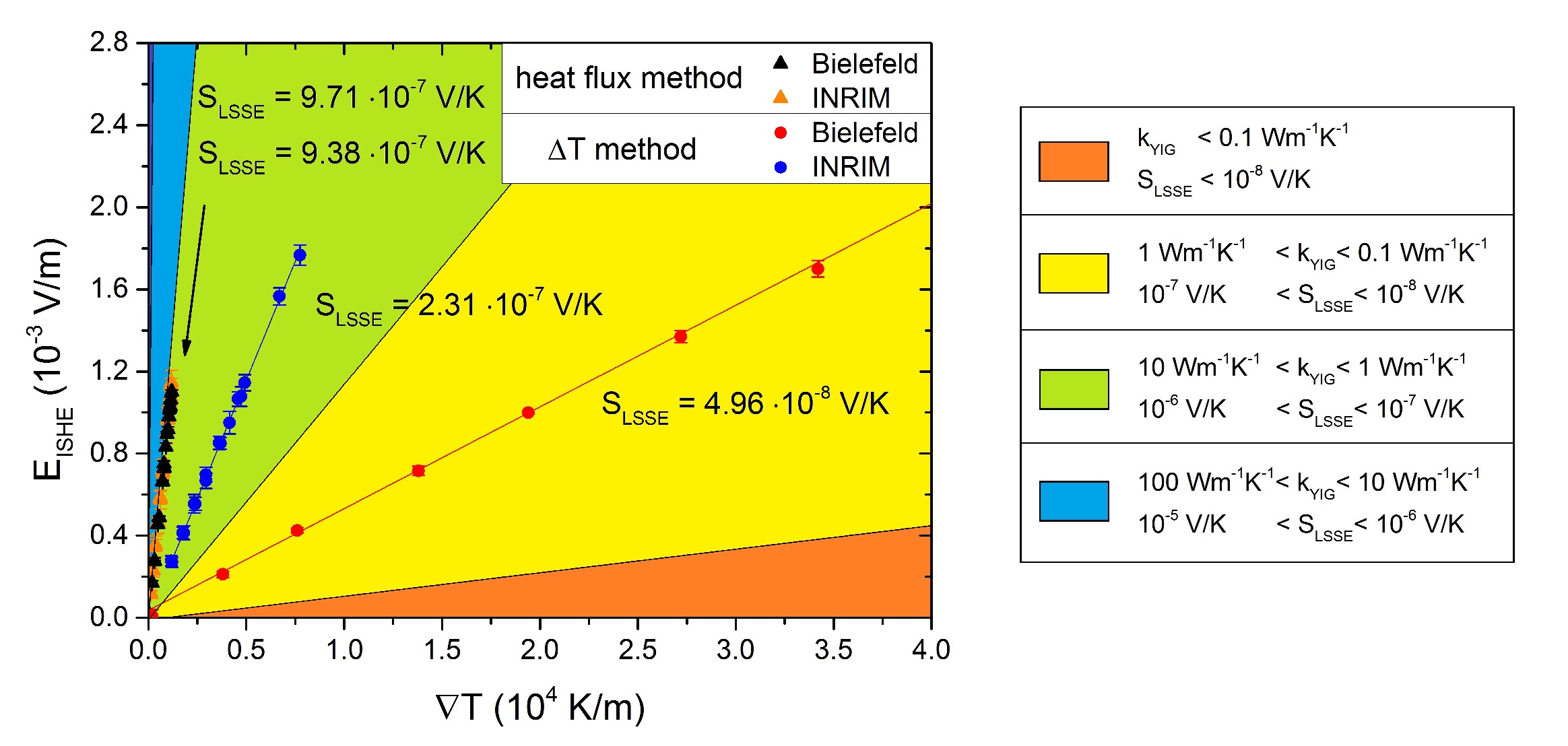}
	\caption{\label{Figure-4}
		LSSE driven electric field as a function of the thermal gradient. Two series (red and blue dots obtained at Bielefeld University and INRIM, respectively) refer to the $\Delta T$ method where the gradients on the x-axes are calculated as $\nabla T=\Delta T/L_{z}$. The series represented by black and orange triangles (obtained at Bielefeld University and INRIM, respectively) refer to the heat flux method and the gradients calculated as $Q\cdot k_{\textmd{YIG}}/A$.
		The color areas represent the ranges of slopes ($S_\textmd{LSSE}$ in V/K units) obtained with the heat flux method considering different ranges of $k_\textmd{YIG}$ for the calculation of the thermal gradients.
	}
\end{figure}

The possibility to convert the heat flux into a thermal gradient allows the comparison of the heat flux method (Figure \ref{Figure-3}(b)) with the $\Delta T$ method (Figure \ref{Figure-3}(a)). This comparison requires to know the thermal conductivity of the YIG $k_\textmd{YIG}$, the active material that constitutes the sample. The measurement of the value of $k_\textmd{YIG}$, as i.e. performed by Euler and coworkers\cite{euler2015thermal} for a thin film, is beyond the scopes of this work.
Instead, it is possible to represent results from $\Delta T$ method together with some possible ranges of results from the heat flux method whose values depend on the range of thermal conductivities $k_\textmd{YIG}$ of the YIG film. This allows to express the intrinsic property of the LSSE material, which is LSSE coefficient $S_\textmd{LSSE}$ in V/K, from measurements obtained with the heat flux method for different values of $k_\textmd{YIG}$.
For the sample under test, an active layer with $k_\textmd{YIG}$ between $1$ and $10$ Wm$^{-1}$K$^{-1}$, which is the most realistic range of values ($6.63$ Wm$^{-1}$K$^{-1}$ for bulk\cite{hofmeister2006} and $8.5$ Wm$^{-1}$K$^{-1}$ for thin films\cite{euler2015thermal}), gives a LSSE coefficient $S_\textmd{LSSE}$ between $10^{-7}$ and $10^{-6}$ V/K.
The green region of Figure \ref{Figure-4} represents these values, while other color regions express the values that $S_\textmd{LSSE}$ would represent if the LSSE material had other thermal conductivity values (see color legend in Figure \ref{Figure-4}).

As an example of this comparison, we choose a value for $k_\textmd{YIG}$ as equal to $8.5$ Wm$^{-1}$K$^{-1}$: this value is obtained at room temperature from films of different thicknesses ($6.7$ $\mu$m, $2.1$ $\mu$m and 190 nm) by Euler and coworkers\cite{euler2015thermal}.
The thermal gradients measured with heat flux method and calculated with $k_\textmd{YIG}=8.5$ Wm$^{-1}$K$^{-1}$ give values of $S_\textmd{LSSE}$ coefficient equal to $S_\textmd{LSSE}=(9.379\pm0.062)\cdot10^{-7}$ V/K and $(9.705\pm0.053)\cdot10^{-7}$ V/K for Bielefeld University and INRIM, respectively.
Instead, for what concerns the $\Delta T$ method, the LSSE coefficients $S_\textmd{LSSE}$ measured by the two groups deviates by a factor of $4.6$ as shown in Figure \ref{Figure-3}(a) and exhibit a significant underestimation of its value when compared to the heat flux method.

\section*{Discussion}

There is an evident mismatch between the values of LSSE coefficient obtained in the two laboratories with the $\Delta T$ method, as shown in Figure \ref{Figure-3}(a). These results imply that the estimation of the temperature profile represented in Figure \ref{Figure-1}(b) does not approximate well the real one. The thermal resistance of the contacts leads to a substantial temperature drop so that the temperature difference across the sample is largely overestimated by this method.
In order to obtain a good estimate of the real temperature difference across the sample with this method, the thermal resistance of the contacts should be negligible with respect to the one of the sample.
If we consider  the thermal conductivity of the substrate $k_\textmd{YAG}=14$ Wm$^{-1}$K$^{-1}$ as an approximation of the thermal conductivity of the whole sample which has lateral dimension of 0.5$\times$2.5$\times$5 mm$^{3}$, the value of its thermal resistance is about $3$ K/W. Instead, the thermal resistance of the contacts can reach values ten times larger and, in some specific configurations, can have a random fluctuation of this value larger than the thermal resistance of the sample, as was shown in a preliminary study on the reproducibility of LSSE measurements\cite{sola2015evaluation}. 
In this work we study the reproducibility of the thermal contact by performing two series of temperature difference as function of the heat flux across a LSSE sample. In two series, the thermal contact is fabricated with thermal grease and we show that the results depend strongly on the quantity and the homogeneity of the thermal grease together with the pressure exerted for the clamping of the LSSE sample.
One possibility to overcome these problems and correctly determine the temperature difference across the magnetic layer was proposed by Uchida et al.\cite{uchida2014quantitative}. In this work, the temperature at both sides of a thick YIG film is measured via the resistance characteristics of two Pt layers deposited directly on the YIG film. In fact, the LSSE coefficients obtained by this method differ largely from the ones published before\cite{uchida2014longitudinal}. However, this method is only applicable to a sample in which is possible to deposit a Pt film on each surface, like for a thick sample.

The second comparison, reported in Figure \ref{Figure-3}(b), regards the heat flux method. The structural features of the systems in the two laboratories are the same and the calibration of the Peltier sensors has been performed according to the same procedure described in the Methods section.
The measurement of the electric field exhibits the same uncertainties as described for the $\Delta T$ setup, while the response of the Peltier sensor in terms of heat flux is not affected by the thermal resistance of the contacts. Therefore, it is possible to eliminate the systematic error due to this thermal resistance. While using the heat flux method, the uncertainty between two sets of measurements from the same laboratory is of the same order as the uncertainty between two measurement sets from the two laboratories.
Systematic errors related to heat leakage from the thermal circuit can be considered as very small (Figure \ref{Figure-1}(c)), and it is easier to control the heat leakages in a thermal circuit than to control the value of the thermal resistance of the contacts between the temperature sensors and the sample.

For what concerns the heat flux method, it is necessary to have information about the thermal conductivity of the material in order to obtain the intrinsic coefficient $S_\textmd{LSSE}$. The value of $k_\textmd{YIG}$ is not easily accessible by an experimental analysis, especially for a thin film and the role of this quantity in the evaluation of $S_\textmd{LSSE}$ from the heat flux measurement is represented by different color areas in Figure \ref{Figure-4}. However, a variation of one order of magnitude for $k_\textmd{YIG}$ leads to a variation of the corresponding coefficients $S_\textmd{LSSE}$ which is comparable to the variation experimentally observed between the $\Delta T$ measurements. Indeed, the determination of thermal conductivity $k_\textmd{YIG}$ is essential for the heat flux method but still the expression of its value with reasonably large uncertainty allows better accuracy of $S_\textmd{LSSE}$ with respect to the $\Delta T$ method.
By approximating the thermal conductivity of the sample under test with the value reported in literature for a thin film $k_\textmd{YIG}=8.5$ Wm$^{-1}$K$^{-1}$\cite{euler2015thermal}, it is possible to observe that the $S_\textmd{LSSE}$ obtained by the $\Delta T$ method tends to be underestimated, as reported in the comparison of the values of $S_\textmd{LSSE}$ obtained with the heat flux method and the $\Delta T$ method in Figure \ref{Figure-4}. This is due to the fact that the temperature gradient evaluated by the heat flux method is concentrated across the sample only, while with the $\Delta T$ method the gradient includes contributions of the thermal contacts between sample and thermal baths.

In summary, we have examined the reproducibility of the LSSE measurements with two experimental methods namely the $\Delta T$ and the heat flux method. We found that the characteristics of a LSSE sample can be measured reproducibly and with a low uncertainty only by using an approach based on the heat flux method. Future work will be directed to the quantitative characterization of SSE materials using this approach. The comparison was performed by INRIM and Bielefeld University and advantages and disadvantages that we pointed out for these two methods are summarized in Table \ref{tab:1}.

\begin{table}[ht]
	\centering
	\begin{tabular}{|l|l|l|}
		\hline
		$S_{\textrm{LSSE}}=-E_{\textrm{ISHE}}/\nabla T$ & $\Delta T$ method $\rightarrow$ $\nabla T=\Delta T/L_{z}$ & Heat flux method $\rightarrow$ $\nabla T=(Q/A)/k_{YIG}$\\
		\hline
		$\nabla T$ & Measurement of $\Delta T$ & Measurement of $(Q/A)$ \\
		\hline
		Reproducibility & Bad: tricky control of thermal contacts  & Good: high control on heat leakages \\
		\hline
		Accuracy & Limited by the knowledge about thermal resistances & Limited by the knowledge about $k_\textmd{YIG}$ \\
		\hline
		\hline
		Thin film sample & External thermometers required & Tricky thermal conductivity measurement \\
		\hline
		Bulk sample & Optimal for a double Pt film $\Delta T$ measurement & Optimal for thermal conductivity measurement \\
		\hline
	\end{tabular}
	\caption{\label{tab:1}Summary comparison: heat flux vs. temperature difference method.}
\end{table}

\section*{Methods}

\subsection*{Sample}
The LSSE sample is a 60 nm thick YIG film that was deposited on 0.5 mm thick yttrium aluminium garnet (Y$_3$Al$_5$O$_12$) (111)-oriented single crystal substrates with 5 $\times$ 2.5 mm$^2$ in dimension by pulsed laser deposition from a stoichiometric polycrystalline target. The KrF excimer laser had a wavelength of 248 nm, a repetition rate of 10 Hz and an energy density of 2 J/cm$^{2}$.
A thin film (3 nm) of Pt was deposited on the YIG surface and we sputtered two 100 nm thick gold electrode strips at the edges of the sample top surface. These electrodes guarantee the same ohmic contact between the Pt film and the electrical connections used in the two laboratories; these were 30 $\mu$m diameter Al wires connected to the sample by wire-bonding in Bielefeld and 40 $\mu$m diameter Pt wires connected with silver paste at INRIM.
In the absence of patterned electrodes the measured voltage may depend on the characteristic of the contact, i.e. the distance between the contacts and the dimensions of the bonding or silver glue spot.

\begin{figure}[ht]
	\centering
	\includegraphics[width=6cm]{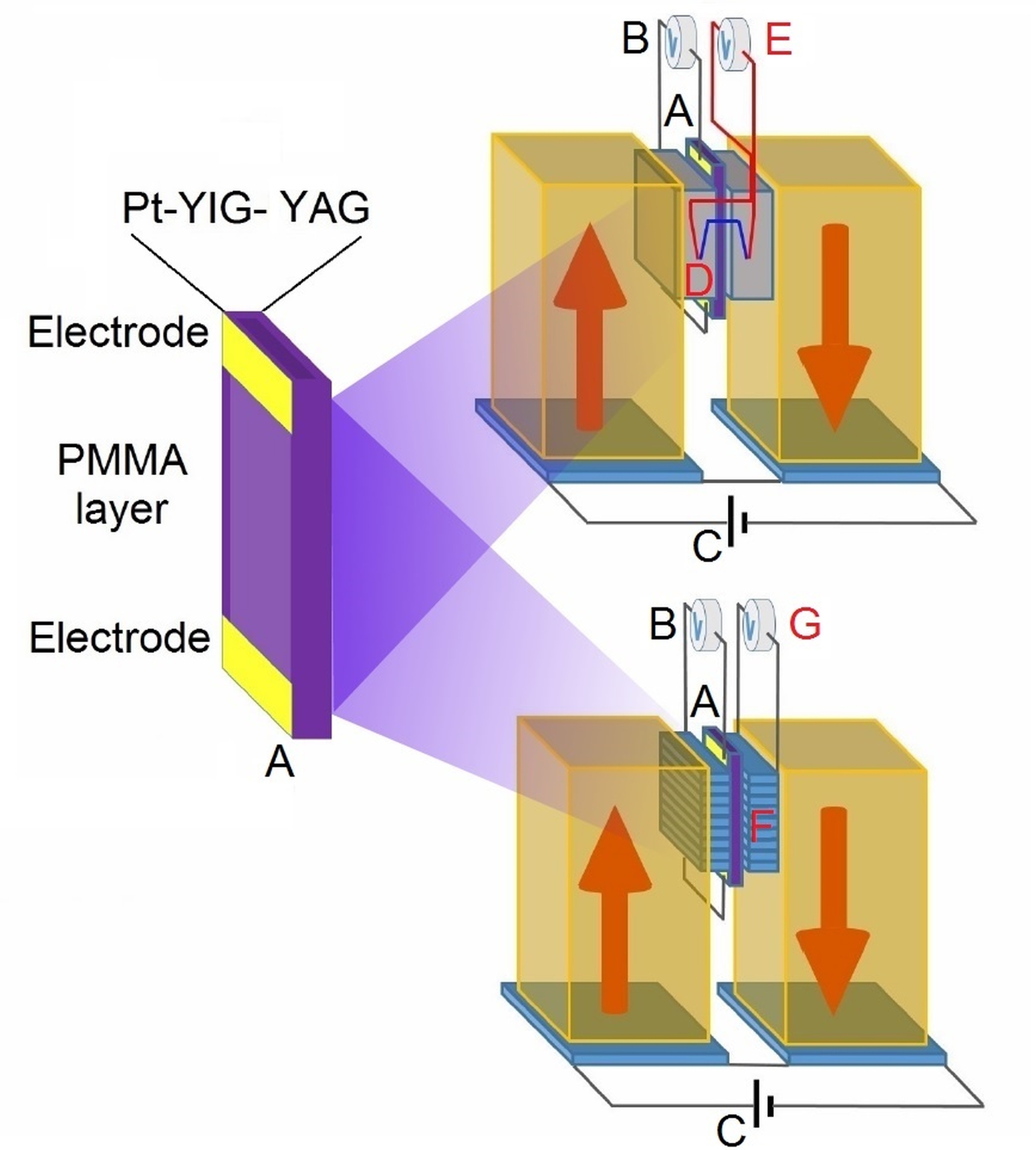}
	\caption{\label{Figure-5}
		Schematic illustration of the $\Delta T$ method to characterize the LSSE sample (upper sketch) and scheme of the heat flux method (lower sketch). (A) LSSE sample (B) nano-voltmeter for the measurement of ISHE voltage (C) Peltier elements employed to the heat current generation (red arrows inside the brass blocks). (Top panel) $\Delta T$ configuration featured by (D) thermocouples and (E) nano-voltmeter to monitor the thermocouples output. (Bottom panel) heat flux measurement configuration with (F) calibrated Peltier sensors and (G) nano-voltmeter for acquiring the heat flux sensor output.
	}
\end{figure}

Moreover, we spin-coated the Pt surface of the sample with PMMA as a protection layer. This is required by the large number of measurement cycles with different systems which can cause deterioration of the Pt film. This preparation step increases the contact thermal resistance and adds a constant value of temperature drop to the measurement in conventional $\Delta T$ method configuration while it does not affect the measurement of the heat flux, since the heat flux method is independent from the thermal contact resistances.
The heat flux and the $\Delta T$ measurement systems are sketched in Figure \ref{Figure-5}.

\subsection*{Measurement setup}

The $\Delta T$ method setup (upper sketch of Figure \ref{Figure-5}) was equipped with two thermocouples (K-type in Bielefeld and T-type in INRIM), a heat current actuator which can be either a Peltier element or a Joule heater and some thermal junctions that allow to insert the sample in the circuit. Thermal junctions are used for both the heat flux and the $\Delta T$ method because it is necessary both to cover the whole surface of the sample and to adapt it to the geometry of the thermal bath and the heat flux sensor. Since these thermal junctions are directly in contact with the LSSE sample, they have to be both electrical insulators and heat conductors; sapphire (Al$_2$O$_3$) was used in Bielefeld while aluminum nitride (AlN) was used at INRIM.
We used thermal junctions of the proper geometry in order to counteract any possible effect of the difference between their values of thermal conductivities. The sapphire slab, whose nominal thermal conductivity is equal to (23 - 25 Wm$^{-1}$K$^{-1}$) is 0.5 mm thick, while the aluminum nitride slab with thermal conductivity equal to (140 - 180 Wm$^{-1}$K$^{-1}$) is 3 mm thick.
The differences in temperature drop across the chosen AIN and sapphire junctions are smaller than 0.05 K for the quantities of heat involved in this experiment that are below $10^{4}$ Wm$^{-2}$.
The temperature difference sensed by the thermocouples as a voltage and the ISHE voltage at the Pt edges are both electrically probed by means of two Keithley 2182 nanovoltmeters. As reported in Figure \ref{Figure-1}(a), the two thermocouples are placed in a region that is supposed to be at thermal equilibrium (hot and cold baths).

The heat flux method measurement system is depicted in the lower sketch of Figure \ref{Figure-5} and includes two Peltier elements which are working as heat flux actuators in order to sustain a heat current loop through all the other elements placed in series: the aluminum nitride and brass connecting elements, the calibrated Peltier sensors and the LSSE sample. In order to avoid undesirable heat flux leakage, the system is kept under vacuum (1.6$\times10^{-4}$ mbar obtained by a turbomolecular pump). Vacuum is maintained both during the Peltier calibration procedure and during the LSSE measurements. Since all electrical connections also produce heat flux leakage the dimensions of the electrical wires of the sensors are 150 $\mu$m diameter and 4 mm of length; this guarantees a negligible heat leakage, when compared to the sample-sensor series circuit heat conductance. It is not possible to control the heat leakage due to infrared radiation in this setup; however, the temperatures involved are low enough to consider the heat losses negligible according to the Stefan-Boltzmann law.

\subsection*{Calibration}

\begin{figure}[ht]
	\centering
	\includegraphics[width=11cm]{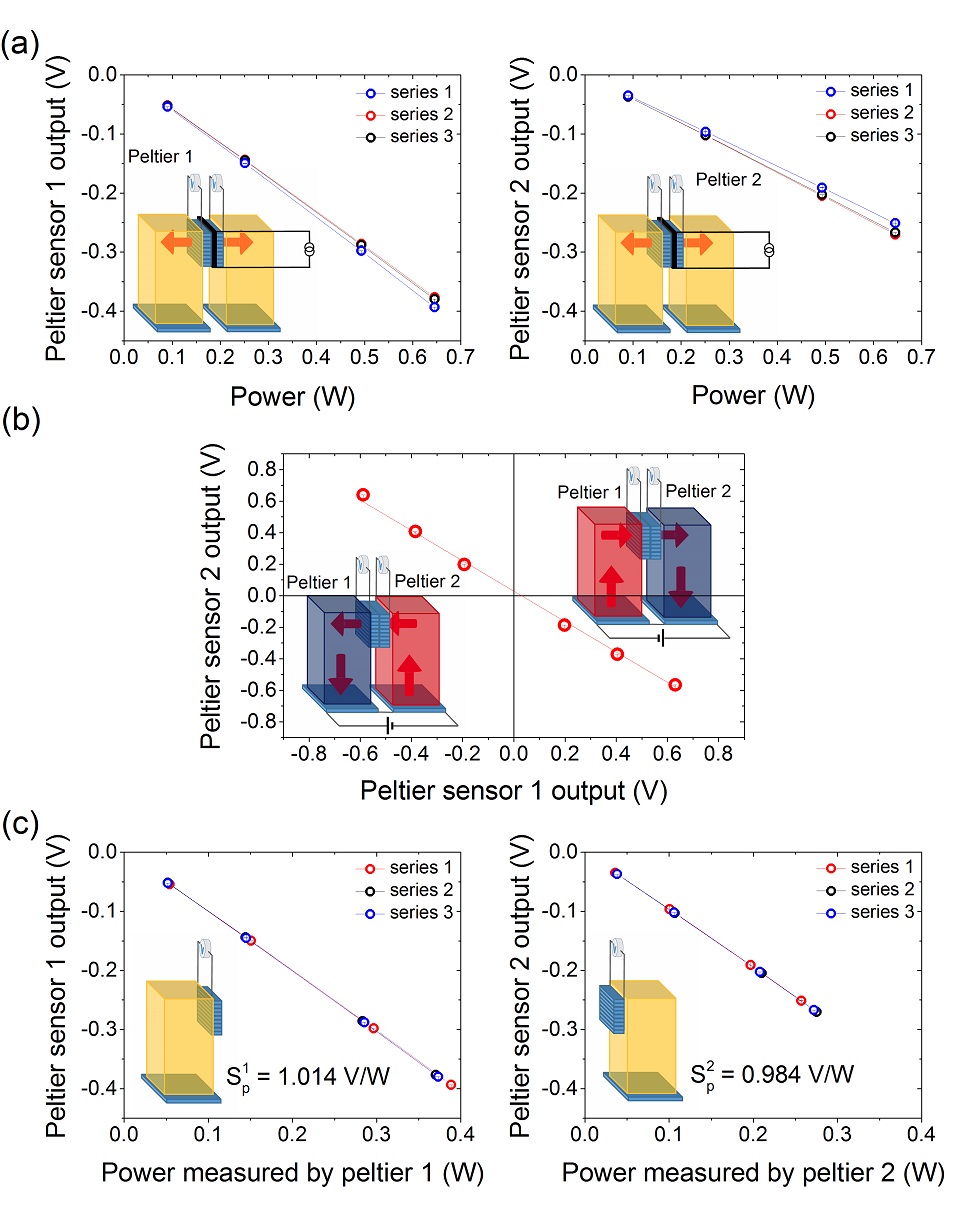}
	\caption{\label{Figure-6}
		Calibration procedure for the Peltier heat flux sensors. (a) 3 sets of voltage responses for each Peltier sensors as a function of the whole power produced by the Joule heater. (b) Voltage output from the first sensor as a function of the voltage output of the second sensor. (c) Voltage responses for each Peltier sensors as a function of the amount of power which diffuses through it; these curves are the characteristics of each Peltier sensor in V/W.
	}
\end{figure}

While using  the heat flux method, a calibration procedure is required to obtain the voltage as a function of the power characteristic of the Peltier sensors; this procedure is summarized in Figure \ref{Figure-6}. 

In the first step shown in Figure \ref{Figure-6}(a), we use a 100 $\Omega$ SMD resistor as Joule heater clamped between the two Peltier sensors in the same position where we place the LSSE sample. We apply a predefined current into the heater using a Keithley 2601 source meter that also measures the power dissipation. Small resistance variations of the SMD due to thermal cycles are not affecting the calibration thanks to the direct measurement of power. This first step gives information about the sum of the responses of the two Peltier sensors as a function of the power produced by the Joule heater, but it is not possible to determine the proportion in which the heat is distributed among the two Peltier sensors.

This information is obtained from a second calibration step, which requires the presence of the same amount of heat flowing through the interconnected Peltier sensors. This second step is shown in Figure \ref{Figure-6}(b), where we plot the voltage response of each Peltier sensor as a function of the voltage response of the other one, under the hypothesis that the same heat flux is flowing through both sensors. The behavior of the two Peltier sensor can be expressed by $U_{2}=F\cdot U_{1}$, where $U_{1}$ and $U_{2}$ are the voltage responses of Peltier 1 and Peltier 2, respectively, and $F$ is the slope of the curve in Figure \ref{Figure-6}(b). 
We proved the absence of heat leakages during this process with the following method: we have purposely varied the thermal resistance at the interface between the two Peltier sensors by adding and removing subsequently a slab of silicon wafer from this region. The factor $F$ remains the same for different levels of thermal resistance produced between the two Peltier sensors, giving evidence of zero heat leakage during the calibration.
By matching the information from Figure \ref{Figure-6}(a) and \ref{Figure-6}(b), it is possible to obtain the sensitivities of the two Peltier sensors from the quantity of power that stimulates their voltage responses according to the following expressions: $P_{1}=(P_{tot}\cdot U_{1})/(U_{1}+U_{2}/F)$ and $P_{2}=(P_{tot}\cdot (U_{2}/F))/(U_{1}+U_{2}/F)$, where $P_{1}$ and $P_{2}$ are the amounts of power that is crossing the two sensors Peltier 1 and Peltier 2 and they fulfill the condition $P_{1}+P_{2}=P_{tot}$. The characteristics of the two Peltier sensors in $V/W$ are reported in Figure \ref{Figure-6}(c); these values are $S_{p1}=(1.014\pm0.002)\,V/W$ and $S_{p2}=(0.984\pm0.002)\,V/W$. We can neglect the temperature dependence of the Seebeck coefficient because both the calibration and the measurements are performed around room temperature.
It is possible to perform the LSSE measurements with a single Peltier sensor, under the hypothesis of a negligible heat leakage along the circuit formed by the LSSE sample and the Peltier sensor. However, since the calibration process provides the characteristics of two sensors, we checked a LSSE characterization as function of heat flux measured by both Peltier sensors. For low levels of heat flux (below $6\cdot10^{3}$ Wm$^{-2}$), the two Peltier sensors monitor the same values, while their output diverges by $3.5 \%$ for high levels of heat flux ($10^{4}$ Wm$^{-2}$). This is a sign of a slight rise of heat leakage through the electrically connected sample.

\providecommand{\noopsort}[1]{}\providecommand{\singleletter}[1]{#1}%

\section*{Acknowledgements}

We thank the DFG (SPP 1538) and the EMRP JRP EXL04 SpinCal for financial support. The EMRP is jointly funded by the EMRP participating countries within EURAMET and the EU. TK and GR received funding from the DFG priority programme SpinCaT (Ku 3271/1-1 and Re 1052/24-2).
We thank Sibylle Meyer from Walther-Meissner-Institut, Garching, Germany for the sample preparation.

\section*{Author contributions statement}

A.S., P.B., D.M. and T.K. performed the experimental work. V.B. and M.K. proposed the measurement technique. A.S. wrote the main manuscript and prepared all the figures. G.R., and M.P. proposed the research approach. 
All authors discussed the experimental details and reviewed the manuscript.

\section*{Additional information}

\textbf{Competing financial interests} The authors declare no competing financial interests.



\end{document}